\documentclass[aps,prd,twocolumn,groupedaddress]{revtex4-2}

\usepackage{graphicx}
\usepackage{amsmath,amssymb,latexsym,slashed,multirow,bm,hyperref}

\def\re{{\rm Re\,}}
\def\tr{{\rm Tr\,}}

\def\lmatrix{\left(\begin{array}}
\def\rmatrix{\end{array}\right)}
\def\bea{\begin{eqnarray}}
\def\eea{\end{eqnarray}}
\def\nn{\nonumber}
\def\msbar{\overline{\rm MS\kern-0.5pt}\kern0.5pt}
\def\gsim{\mathrel{\rlap{\lower4pt\hbox{\hskip1pt$\sim$}}\raise1pt\hbox{$>$}}}
\def\lsim{\mathrel{\rlap{\lower4pt\hbox{\hskip1pt$\sim$}}\raise1pt\hbox{$<$}}}
\def\rho{\varrho}

\begin{document}

\title{QCD on the 16-cell honeycomb}

\author{Sandor D. Katz}
\email[]{katz@bodri.elte.hu}
\affiliation{Eotvos Lorand University, Institute of Physics and Astronomy, Department of Theoretical
Physics, Budapest, Hungary}

\author{Daniel Nogradi}
\email[]{nogradi@bodri.elte.hu}
\affiliation{Eotvos Lorand University, Institute of Physics and Astronomy, Department of Theoretical
Physics, Budapest, Hungary}

\date{\today}

\begin{abstract}
We formulate QCD discretized
    on the four dimensional 16-cell honeycomb. The advantage is a higher degree of rotational symmetry
    as compared to a traditional cubic lattice leading to much smaller cut-off effects. 
    We demonstrate in quenched QCD,
    through both gluonic and fermionic observables, that the scaling
    properties are indeed superior to the cubic lattice and much larger lattice spacings are sufficient for 
    controlled continuum extrapolations. Chiral and topological properties also show remarkable improvement.
\end{abstract}

\maketitle

\section{Introduction\label{introduction}}

The most successful non-perturbative formulation of QCD uses a four-dimensional Euclidean spacetime lattice. Large-scale state of the art lattice simulations face two major challenges: the chiral and continuum extrapolations, both of which are sensitive to the chosen discretization. 
To preserve as much of the continuum spacetime symmetry as possible, one can base the discretization on regular spacetime tessellations -- that is, tilings of Euclidean space by regular polytopes. In two dimensions, three regular tessellations exist (triangular, square, and hexagonal), while in three dimensions only the cubic tessellation is possible. In four dimensions, however, there are again three options: the 16-cell honeycomb, the tesseractic honeycomb, and the 24-cell honeycomb.

Large-scale QCD simulations have so far relied exclusively on the cubic lattice defined by the vertices of the tesseractic honeycomb. The other two regular four-dimensional tessellations, however, possess larger remnant rotational symmetry. Specifically, the symmetry group of the cubic lattice has 384 elements, whereas the symmetry groups of the vertex sets of both the 16-cell and 24-cell honeycombs contain 1152 elements.

In this work, we investigate lattice QCD formulated on the vertices of the 16-cell honeycomb. Interestingly, this lattice corresponds to the densest sphere packing in four-dimensional Euclidean space: each site has 24 nearest neighbors, equal to the four-dimensional kissing number. We find that cutoff effects are substantially reduced and that lattice spacings lying outside the scaling regime of the cubic formulation fall comfortably within it on this lattice. Furthermore, chiral properties are also superior to the cubic discretization.

The first hints that cut-off effects are much smaller using the 16-cell honeycomb rather than the cubic lattice come from free
fields. For instance, the pressure of massless free fermions at finite temperature, if renormalized
suitably, has ${\cal O}(a^2)$ cut-off effects on a cubic lattice but only ${\cal O}(a^4)$ for the 16-cell honeycomb. Beyond free fields, we study the cut-off effects of the topological susceptibility in $SU(3)$ gauge theory and show 
that even in the interacting case any ${\cal O}(a^2)$ cut-off effects are small enough to not be visible in our data.

Turning to the naive discretization of fermions it should be noted that the 16-cell honeycomb was investigated before 
\cite{Celmaster:1982ht,Celmaster:1983jq,Celmaster:1983vy,Celmaster:1983hs,
Celmaster:1984rf,Celmaster:1985cv,Celmaster:1986eb}
and the dispersion relation of some of the doublers were shown to 
break Lorentz invariance. This
problem is eliminated, however, by adding a Wilson term to the action which pushes the mass scales of the
doublers to the cut-off scale as usual. The physical mode remains massless and its dispersion relation Lorentz
invariant, leading to a physical spectrum free from any Lorentz breaking in the continuum. We show, in
quenched QCD, that the Dirac spectrum and the topological properties of low modes is quantifiably closer
to the continuum than the corresponding quantities on a cubic lattice, at fixed lattice spacing.

It should be noted that the same lattice and its sub-lattices 
were also investigated in Higgs models in the context of triviality
in the past \cite{Neuberger:1987kt, Bhanot:1990zd, Bhanot:1990ai}. Exactly because scalar QFT is trivial,
the strict continuum limit cannot be addressed of course.

Our work is the first to report on the
systematics of the continuum extrapolation of $SU(3)$ gauge theory and the properties of quenched
QCD on this lattice of highest possible symmetry in 4 dimensions. 

\section{The 16-cell honeycomb\label{16cell}}

The vertex set of the 16-cell honeycomb is known by multiple names: the body-centered cubic or
tesseractic (BCH or BCT) lattice, or the $D_4, D_4^*$, or $F_4$ lattice. These names have either
crystallographic or group theoretical origin, while the 16-cell honeycomb is a purely geometric concept
which also incorporates the edges, faces and cells of the tessellation, not only its vertices. The
geometric description is the following. The 16-cell honeycomb is one of the three regular space-filling
tessellations in 4 dimensions, the other two being the cubic and the 24-cell honeycomb. The latter
case can be obtained from the 16-cell honeycomb by removing some of the vertices and edges while
retaining the same symmetry.
An analogy can be drawn with the familiar regular space-filling tesselations in 
2 dimensions. There are again three, triangular, cubic, hexagonal. The triangular, which is analogous
to our 16-cell honeycomb in 4 dimensions, is more symmetric than the cubic. 
The third one, the hexagonal, which is analogous to the 24-cell
honeycomb in 4 dimensions, has the same amount of symmetry as the triangular, but some vertices 
and edges are removed.

In order to emphasize the geometric origin of our discretization, in this paper we retain the name as 16-cell honeycomb.

The lattice sites in 4-dimensions can be defined for instance by requiring that
all four components are integers and their sum is even, in lattice units ($D_4$ lattice). Alternatively, one may define
it by having a usual 4-dimensional cubic lattice and adding the centers of each hypercube (BCH lattice). Any
lattice can be defined by a set of basis vectors and lattice sites are given by integer linear
combinations. The set of basis vectors is of course not unique: if these are arranged as columns 
into a $4 \times 4$ matrix $e$ then transformations of the type $e \to \lambda R\, e\, K^{-1}$ are immaterial, where
$\lambda \in {\mathbb R}$ non-zero, $R \in O(4)$ and $K \in GL(4,{\mathbb Z})$. Here $\lambda$
corresponds to an overall rescaling, $R$ to an overall rotation of the basis
vectors while $K$ simply relabels the integers. Two frequently used choices of $e$ basis matrices for the 16-cell honeycomb are the following,
\bea
\label{e}
\lmatrix{cccc} 1 & 0 & 0 & 0 \\ -1 & 1 & 0 & 0 \\ 0 & -1 & 1 & 1 \\ 0 & 0 & -1 & 1 \rmatrix \quad
\lmatrix{cccc} 1 & 0 & 0 & 1/2 \\ 0 & 1 & 0 & 1/2 \\ 0 & 0 & 1 & 1/2 \\ 0 & 0 & 0 & 1/2 \rmatrix,
\eea
and depending on context one is more useful than the other.
Each lattice site thus has 24 nearest neighbors, 12 of these in the positive and 12 in the negative
direction. This enlarged set of nearest neighbors is the main reason for the enlarged rotatational
symmetry. 

Concretely, the second choice in (\ref{e}), used by us also, gives the lattice sites of the 16-cell honeycomb as
follows. The total set of sites is the union of
lattice sites of a cubic lattice and another one shifted by 
$(1/2,1/2,1/2,1/2)$. This means the full lattice consists of
two sublattices and the sites on each can be labelled by four integers $(n_1,n_2,n_3,n_4)$. The 
coordinates on the first sublattice are directly these $(n_1,n_2,n_3,n_4)$ while the coordinates
on the second sublattice are $(n_1 + 1/2, n_2 + 1/2, n_3 + 1/2, n_4 + 1/2)$.

Every lattice site is equivalent, of course, and the origin has the following 24
nearest neighbors: there are 8 usual neighbors on its cubic sublattice and there
are 16 more on the shifted sublattice. The number $16 = 2^4$ comes from the number of ways the
origin can be shifted by $(\pm 1/2, \pm 1/2, \pm 1/2, \pm 1/2)$ onto the shifted
sublattice.

Once the lattice becomes finite and (anti)periodic boundary conditions are imposed, the choice of
basis will matter, because they will set the directions in which there is (anti)periodicity. At finite
temperature this choice clearly matters in the temporal direction. In the present work the second choice in
(\ref{e}) is used in the simulations. We fix the boundary conditions such that they correspond to an
orthogonal torus: a periodicity of $N_s$ sites is imposed in the $(1000),(0100),(0010)$ directions while
(anti)periodicity of $N_t$ sites is assumed in the $(0001)$ direction. Note that
the boundary conditions break the full rotational symmetry but this has little impact on the observed scaling properties, as was also noted in~\cite{Celmaster:1983vy}.

Using this setup, for a given physical volume and lattice spacing our lattice contains twice as many sites as the cubic one. Since each site has three times as many nearest neighbors, the naive computational cost of the 16-cell discretization is 6 times that of the cubic one.

For more details on numerous aspects of the 16-cell honeycomb see \cite{conwaysloane}.

\begin{figure}
    \includegraphics{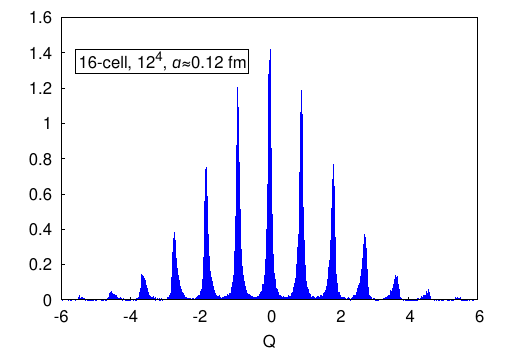}
    \caption{Distribution of the topological charge on the
    16-cell honeycomb. For clarity we do not show the corresponding histogram with the cubic lattice, but as can be expected, the $a=0.12$~fm lattice spacing is too large for clear peaks in that
    case.}
    \label{qhist}
\end{figure}

\section{Gluons\label{gluons}}

Gauge links can naturally be defined for each link of the lattice, just as in the cubic case. An
elementary closed loop is a triangle, however, and so the basic building block of the gauge action will
contain sums over triangles. At each site of the lattice there are 192 oriented triangles emanating from that
site, half of them positively oriented and half of them negatively. Since a triangle has 3 corners, in
order to avoid over counting, we need to sum over 32 triangles for each site, leading to the $SU(N)$ gauge
action,
\bea
\label{action}
S = \frac{\beta}{2} \sum_x \sum_{i=1}^{32} \left( 1 - \frac{1}{N} \re\tr P_i(x) \right)
\eea
where $P_i(x)$ is the product of the 3 links in the given triangle. The overall normalization ensures
that $\beta = 2N/g_0^2$ with bare coupling $g_0$ just as with the cubic action.

We report on the continuum limit and scaling property of the topological susceptibility, 
hence let us define the topological charge from basic triangles. A
symmetric definition is adopted akin to the clover-type definition on the cubic
lattice.  With the 16-cell honeycomb there are 6 basic triangles with a common
corner in a given plane, instead of the 4 basic plaquettes with the cubic
lattice.  These 6 basic triangles form a hexagon and there are 16 hexagons at a
given site. From these 16 hexagons the field strength tensor components are
defined in a symmetric manner, exactly following the construction of the cubic
lattice but replacing plaquettes with triangles, from which the topological
charge density as well as the total topological charge follow.

We use the gradient flow to determine the lattice spacing as described in Ref.~\cite{Luscher:2010iy}.
The flow evolves the fields towards the minima of the classical action in fictitious time $t$. The dimensionful parameter $t_0$ is defined as $\left.\frac{1}{2}t^2F_{\mu\nu}(t)F_{\mu\nu}(t)\right|_{t=t_0}=0.3$. Converting the scale to physical units is ambiguous in pure gauge theory, in the following we use $\sqrt{8t_0} = 0.47$~fm.

The gradient flow can also be used to renormalize the topological susceptibility by evaluating the topological charge on configurations evolved to a physical scale. We evaluate the topological charge at $t=t_0$. 

The discretized gradient flow follows straightforwardly from the action (\ref{action}), one only needs to take
into account the non unit area of a basic triangle. The distribution of the topological charge on the 16-cell honeycomb is shown in figure \ref{qhist} with clear peaks close to 
integer values already at $a=0.12$~fm.

\begin{figure}
    \includegraphics{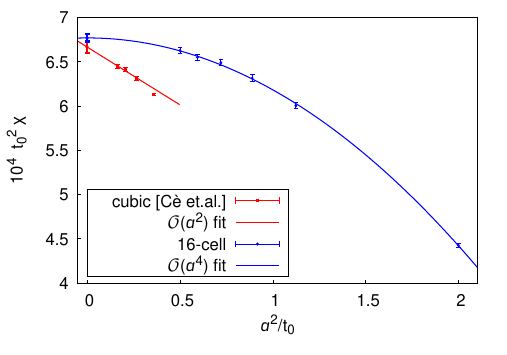}
    \caption{Continuum limit of the topological susceptibility. Solid lines show the extrapolations; see
    text for details.}
    \label{topsusc}
\end{figure}

\begin{table}
\begin{center}
\begin{tabular}{|c|c|c|c|c|}
\hline
    $\beta$ & $L/a$ & $t_0/a^2$ & $L/\sqrt{8t_0}$ & $10^4 \cdot t_0^2\chi$ \\
\hline
\hline
    6.947 &  6 & 0.5005(2) & 2.9986(5) & 4.43(3) \\
\hline
    7.053 &  8 & 0.8901(2) & 2.9979(4) & 6.01(3) \\
\hline
    7.104 &  9 & 1.1260(5) & 2.9987(6) & 6.32(4) \\
\hline
    7.154 & 10 & 1.3899(4) & 2.9989(4) & 6.49(3) \\
\hline
    7.203 & 11 & 1.6845(4) & 2.9965(3) & 6.55(3) \\
\hline
    7.250 & 12 & 2.0022(8) & 2.9984(6) & 6.63(4) \\
\hline
\end{tabular}
\end{center}
\caption{
\label{table}
    Measured $t_0/a^2$ and topological susceptibility in $t_0$ units; data used in figure \ref{topsusc}.}
\end{table}

Once all ingredients are given, figure \ref{topsusc} shows the
continuum limit of the topological susceptibility $\chi$ in $t_0$-units; see table \ref{table} for the 
measured values. The volume is kept fixed also in
$t_0$-units, the bare couplings are tuned to $L / \sqrt{8t_0} = 3$ within 1\% accuracy on lattice volumes
$6^4, 8^4, 9^4, 10^4, 11^4$ and $12^4$. We used standard local overrelaxation and heatbath updates and generated ${\cal O}(50000)$ independent configurations for each lattice.
The results are compared to the corresponding ones on a
cubic lattice from \cite{Ce:2015qha}. While there is clear agreement for the continuum limit, as it should be, the
leading cut-off effects are drastically reduced and appear to follow ${\cal O}(a^4)$ to high
precision. Even though ${\cal O}(a^2)$ corrections are not expected to vanish completely they appear to be highly suppressed. We performed two fits for the continuum extrapolation: one with ${\cal O}(a^2)$ and ${\cal O}(a^4)$ terms and one with only ${\cal O}(a^4)$. The $\chi^2/{\rm d.o.f.}$ values are $0.34$ and $0.30$, respectively and the ${\cal O}(a^2)$ coefficient is consistent with zero in the first case. Its 1$\sigma$ upper bound is an order of magnitude smaller than  the ${\cal O}(a^2)$ coefficient on the cubic lattice. Therefore in the plot we show the ${\cal O}(a^4)$ fit only.
Lattice spacings as large as $a = 0.24$~fm can be used in the fit,
which are definitely outside the leading scaling regime of ${\cal O}(a^2)$ with a cubic lattice. 

\section{Quarks\label{quarks}}

Quark fields are defined at each site as usual. The naive discretization of the free massless Dirac operator,
summing over all neighbors,
\bea
D_0 = \frac{1}{6} \sum_{i=1}^{24} \gamma_\mu n_{\mu i} \nabla_i,
\eea
where $n_{\mu i}$ are the 24 nearest neighbor directions,
leads to Lorentz violating doublers \cite{Celmaster:1983jq}.
Here the overall factor $1/6$ is dictated by the correct
$p^2$ normalization of $D_0^\dagger D_0$ in momentum space.

Let us add a Wilson term with coefficient $r$ which,
just as in the cubic case, pushes the doublers to the cut-off scale leaving a single Lorentz
invariant physical massless flavor in the continuum,
\bea
\label{dirac}
D = D_0 + a \frac{r}{6} \sum_{i=1}^{24} \nabla_i^* \nabla_i,
\eea
where again the factor $1/6$ ensures that in momentum space the Wilson term becomes $a\, r\, p^2/2$ for small
momenta, as usual. The free Wilson propagator has no doublers and has the correct naive continuum limit. Therefore it satisfies all conditions required for the applicability of the lattice power counting theorem~\cite{Reisz:1987da}. We thus expect that lattice perturbation theory works similarly to Wilson fermions on the hypercubic lattice, the continuum limit of all loop integrals with a negative lattice degree of divergence equal to the corresponding continuum Feynman integrals. In particular the general proof of the axial anomaly in Ref.~\cite{Reisz:1999cma} is also applicable ensuring that the Wilson Dirac operator above produces the correct anomaly in the continuum limit. The Lorentz breaking doublers of naive fermions are thus not expected to generate
Lorentz breaking in loop integrals performed over the Brillouin zone.

\begin{figure}
    \includegraphics{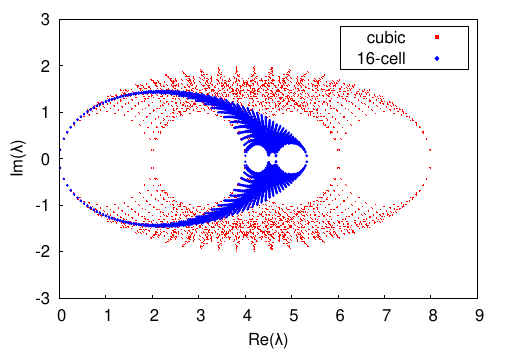}
    \caption{Comparison of the free Wilson-Dirac spectra on lattice volumes $16^4$.}
    \label{freespectrum}
\end{figure}

Figure \ref{freespectrum} shows the free spectrum of the 16-cell honeycomb Wilson-Dirac operator (\ref{dirac})
in comparison with the cubic case. 
A striking observation is obvious: the spectrum
appears to be much closer to a chirally invariant Ginsparg-Wilson-like circle than Wilson fermions on a
cubic lattice, which is very promising.

\begin{figure}
    \includegraphics{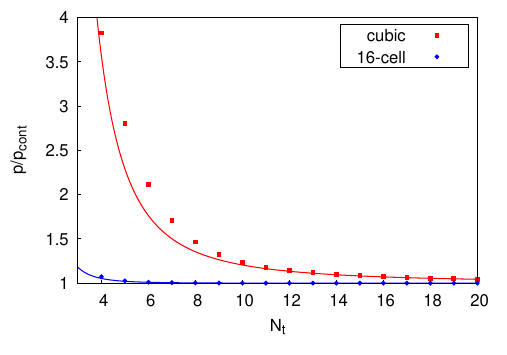}
    \caption{The free pressure as a function of $N_t$ normalized by the continuum values. The solid lines show the analytic ${\cal O}(1/N_t^2)$ and ${\cal O}(1/N_t^4)$ scalings of the two discretizations given by eqn~(\ref{pressure}).}
    \label{pressurefig}
\end{figure}

Another quantity, which in the free case also shows promising improved scaling, is the pressure at finite
temperature. Let us subtract the zero temperature pressure, as usual, in order to have no additive
divergences, and study ${\cal O} = (p(T)-p(0))/T^4$. 
For the free case corrections to the continuum value in infinite spatial volume can be straightforwardly 
computed as a series in $1/N_t^2$, and we obtain,
\bea
\label{pressure}
{\cal O}_{cubic} &=& \frac{7\pi^2}{720} \left( 1 + 
\frac{248}{147}\frac{\pi^2}{N_t^2} + \frac{635}{147}\frac{\pi^4}{N_t^4} \ldots \right)  \\
{\cal O}_{16-cell} &=& \frac{7\pi^2}{720} \left( 1 + 
\frac{127}{980}\frac{\pi^4}{N_t^4} + \frac{73}{4158}\frac{\pi^6}{N_t^6} + \ldots \right). \nn
\eea
The leading corrections in the 16-cell honeycomb case are ${\cal O}(a^4)$ as opposed to ${\cal O}(a^2)$ in the cubic
case, a major improvement.  Figure \ref{pressurefig} shows the scaling behavior at finite $N_t$, in full agreement
with the asymptotic forms (\ref{pressure}). At $N_t = 4$, for instance, the correction to the continuum
is $7\%$ and $283\%$ on the 16-cell and cubic lattices, respectively, while at $N_t = 6$ these are
$1\%$ and $110\%$, respectively. In order to replicate on the cubic lattice the rather small $7\%$ and $1\%$ corrections of the 16-cell
lattice at $N_t = 4$ and $N_t = 6$, one would need $N_t = 16$ and $N_t = 40$. 

The fact that the leading correction is ${\cal O}(a^4)$ can be understood from the form of the 
corrections to the dispersion relation. It is a straightforward expansion in the lattice spacing to
obtain
\bea
D^\dagger D = p^2 - \frac{1}{6} a^2 p^4 + {\cal O}(a^4)
\eea
i.e. at order ${\cal O}(a^2)$ the correction is still Lorentz invariant, the first order where this does
not hold is ${\cal O}(a^4)$. With a cubic lattice, already at order ${\cal O}(a^2)$ we encounter Lorentz
breaking terms. As a consequence, on a 16-cell honeycomb the dispersion relation 
becomes, in Minkowski signature, $E = |\boldsymbol p| + {\cal O}(a^4)$, leading to the absence of the 
${\cal O}(a^2)$ correction in the pressure.

\begin{figure}
    \includegraphics{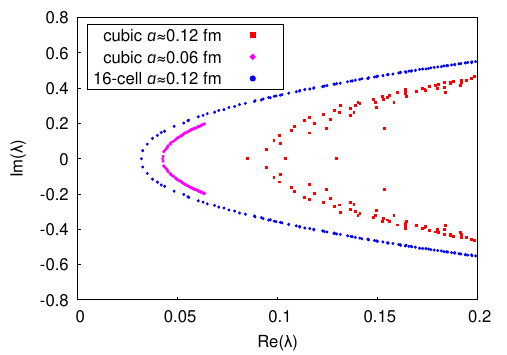}
    \caption{Comparison of quenched Wilson-Dirac spectra on $12^4$ lattice volumes. For the cubic case we also show results on a finer, $24^4$ lattice. Each spectrum corresponds to a single configuration.}
    \label{quenchedspec}
\end{figure}

\begin{figure}
    \includegraphics{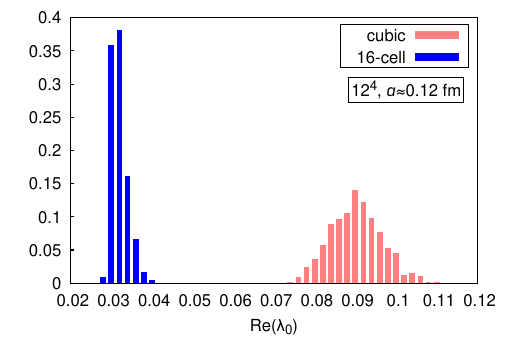}
    \caption{Comparison of the distribution of the lowest Wilson-Dirac eigenvalue.}
    \label{lowmodes}
\end{figure}

Let us turn to the interacting case, inserting gauge links into (\ref{dirac}), and
implementing tree-level clover improvement \cite{Sheikholeslami:1985ij} as well as stout 
smearing \cite{Morningstar:2003gk}. The field strength tensor for clover improvement is obtained at each site from the 16 hexagons, similarly to the gradient flow.
Six steps of stout smearing are applied with smearing parameter $\varrho = 0.05$. In this first work we only study quenched QCD and our focus will be the comparison with the Wilson-Dirac
operator on a cubic lattice using the same improvement, i.e. tree-level clover and stout smearing using the same parameters. In order to have meaningful comparisons the lattice spacings are tuned to $a\approx0.12$~fm for both discretizations. We determined the lowest 300 eigenvalues of the Wilson-Dirac operator on 1215 quenched configurations generated on $12^4$ lattices with both discretizations. In addition, we computed the lowest 100 eigenvalues on a single quenched $24^4$ configuration with $a\approx0.06$~fm

The low-lying quenched Wilson-Dirac spectra evaluated on single configurations are shown in figure \ref{quenchedspec}. 
Two things are immediately clear, first, at the same lattice spacing the additive mass renormalization
will be much smaller on the 16-cell honeycomb and second, the width of the distribution of the real part
of the lowest eigenvalues is also much smaller. The first property results in improved chiral properties
while the second means that smaller quark masses can be simulated at a given lattice spacing, something
with important practical consequences towards the chiral limit. For comparison we also show results on
the cubic lattice at half the lattice spacing and twice the linear lattice size, i.e. the same physical
volume, which still results in larger additive mass renormalization.  For a more detailed look the
distribution of the real part of the lowest eigenvalue is shown in figure \ref{lowmodes}. We observe a
factor of three improvement both for the mean and for the width of the distribution by using the 16-cell
honeycomb.

\begin{figure}
    \includegraphics{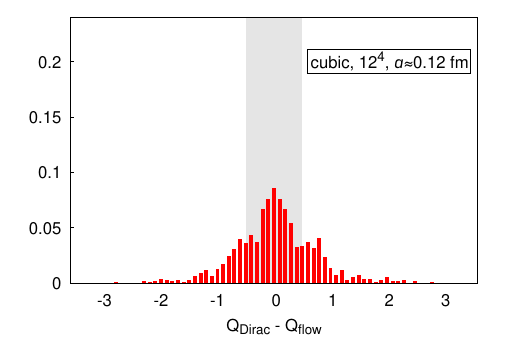}
    \includegraphics{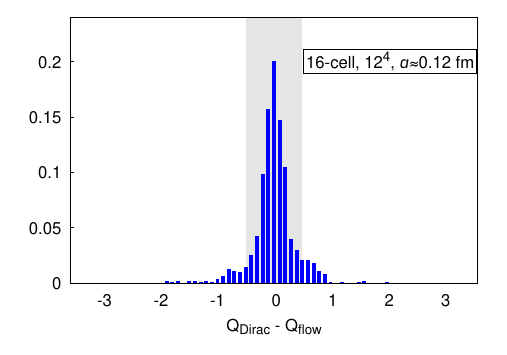} 
    \caption{The distribution of the difference of fermionic and gluonic definitions of the topological
    charge. Top: cubic lattice, bottom: 16-cell honeycomb. The gray bands correspond to the range
    [-0.5:0.5] where a rounding of $Q_{\rm flow}$ would match $Q_{\rm Dirac}$.}
    \label{qdiff}
\end{figure}

The Wilson-Dirac operator has no exact chiral symmetry, nevertheless the low-lying real eigenvalues 
carry important topological information. One may count the number of real low-lying eigenvalues with
non-zero chirality ($n_+$ and $n_-$ for positive and negative chiralities,
respectively)\cite{Hernandez:1997bd} and thus define $Q_{\rm Dirac}=n_+-n_-$ as a topological charge.
Concretely, we take the 20 low-lying real eigenvalues. This choice is of course arbitrary, but varying
it between 10 and 30 does not change the emerging picture.
We compare this fermionic $Q_{\rm Dirac}=n_+-n_-$ to $Q_{\rm flow}$ 
defined using the gradient flow at $t=t_0$ as in the
calculation of the topological susceptibility. These two quantities are of course not necessarily the
same, but towards the continuum limit they are expected to approach each other. The distribution of the
difference $Q_{\rm Dirac} - Q_{\rm flow}$ is shown in figure \ref{qdiff}. The conclusion is the same
as for the quenched Wilson-Dirac spectrum: at the same lattice spacing the 16-cell honeycomb appears to be much closer to the continuum, with practically no configuration showing a difference larger than 1. The index theorem is quite accurately realized already at a lattice spacing $a\approx0.12$~fm. This supports the previous statement that the chiral anomaly is correctly reproduced by our Dirac operator.

\begin{figure}
    \includegraphics{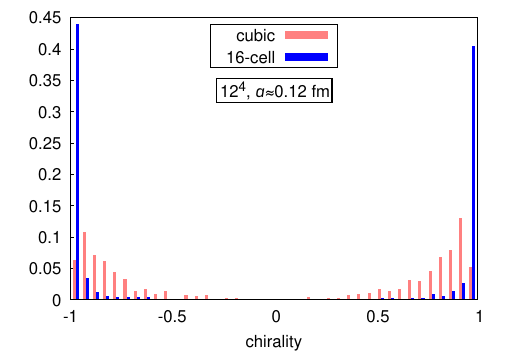}
    \caption{Comparison of distributions of chiralities of the low-lying eigenmodes with the smallest
    real Wilson-Dirac eigenvalues; see text for details.}
    \label{chir}
\end{figure}

Another measure of the topological character of low-lying real modes is how
close or far their chirality is from $\pm 1$. With chirally symmetric
fermion discretizations the zero modes have definite chirality. With Wilson-Dirac fermions this does not
hold but the distribution of the chiralities is a good measure of their topological nature. This is shown
in figure \ref{chir}. The difference between the cubic lattice and the 16-cell honeycomb is again
striking.

Summarizing our findings, it is clear that at comparable lattice spacings the properties of the
Wilson-Dirac operator on the 16-cell honeycomb are much closer to the continuum than on a cubic
lattice. Even though there is an obvious factor 6 increase in computational cost, this is more than compensated
by larger lattice spacings being sufficient for fixed ``continuum-like'' properties, i.e. larger lattice
spacings and smaller lattice volumes are sufficient for controlled and reliable continuum extrapolations.
Furthermore, the smaller additive mass renormalization allows for lower quark masses to be simulated at a given lattice spacing.

\section{Conclusion and outlook\label{conclusion}}

We have investigated lattice QCD discretized on the vertex set of the 16-cell honeycomb, instead of the traditional
cubic one. In several observables ${\cal O}(a^4)$ leading scaling corrections are observed, leading to a
much larger range of lattice spacings from which reliable continuum extrapolations can be performed. 

Our discussion involved gluons and quarks but so far only in the quenched approximation. 
The improved chiral and topological properties of the
Wilson-Dirac operator relative to the cubic case indicates that larger lattice spacings and
consequently smaller lattice volumes can be used for controlled continuum extrapolations. Furthermore,
at a fixed lattice spacing smaller quark masses can be simulated. Both of these properties, better
chiral properties and better continuum scaling lead to significant advantages when considering computational costs. Our results both for gluonic and fermionic observables suggest that on the 16-cell honeycomb about 
twice larger lattice spacings can be used than on the cubic lattice with the same discretization effects. This may lead to as large as an order of magnitude reduction in total computational costs. 


We plan to exploit these features further in fully dynamical QCD simulations in future publications. Even though unquenching effects are usually sizable, the improved continuum and chiral properties of the Dirac operator are expected to be inherited by dynamical simulations and a similar improvement compared to hypercubic discretization could be realistic.

\section{Acknowledgments}
We are grateful for G. Endrődi, M. Giordano, T.G. Kovács and A. Pásztor for useful discussions. This work was supported by NKFIH grants No. TKP2021-NKTA-64, NKKP Excellence No. 151482 and K-147396. All computations were performed on the local GPU cluster at Eotvos University.

\end{document}